\documentclass[preprintnumbers,amsmath,amssymb,floatfix,9pt,prd,onecolumn,
superscriptaddress,nofootinbib]{revtex4}
\usepackage{latexsym,float}
\usepackage{epsfig,color}
\usepackage{epstopdf}
\usepackage{amssymb}
\usepackage{verbatim}
\usepackage{multirow}

\begin{document}
\title{Anisotropic Tolman V Solution by Minimal Gravitational Decoupling Approach}
\author{M. Zubair}
\email{mzubairkk@gmail.com; drmzubair@cuilahore.edu.pk}\affiliation{Department
of Mathematics, COMSATS University Islamabad, Lahore Campus, Pakistan}
\author{Hina Azmat}
\email{hinaazmat0959@gmail.com}\affiliation{Department
of Mathematics, COMSATS University Islamabad, Lahore Campus, Pakistan}

\begin{abstract}
In this paper,  we consider well known Tolman V perfect fluid solution and extend it to its
an anisotropic version using gravitational decoupling by minimal geometric deformation
approach and analyze the behavior of new version of the solution graphically.
The effects of coupling constant on anisotropic factor has been measured and presented
graphically. The matching conditions at the surface of inner and outer geometry have also been discussed. The viability of the
solution has been studied by presenting the physical analysis of the solution.
\end{abstract}

\maketitle


\date{\today}

\section{Introduction}

Gravitational decoupling via minimal geometric deformation (MGD)
approach opens a new horizon for the construction of
anisotropic solutions to Einstein field equations, which is found to be a difficult
task in the presence of non-linear terms there. Although it has
simple features, but it proves itself a very powerful tool which provides a better understanding
of self-gravitating sources with anisotropic backgrounds.
This is a direct, systematic and simple approach that splits
a complex system of equations into two different but simple
set of equations. Here, one corresponds to the usual Einstein
field equations associated with an isotropic background, while the other
is governed by an extra
gravitational source $\Theta_{\alpha\beta}$ which explores the anisotropic configurations.
This novel concept was first suggested by Ovalle \cite{1} in order to obtain the analytic solutions of stellar configurations having
braneworld background. Ovalle and Linares \cite{2} considered isotropic spherical system and worked out the exact solution in the same background
which was found compatible with the Tolman-IV solution.
 Casadio et al. \cite{3} adopted this strategy and derived the exterior solutions of spherical system.

Ovalle and his collaborators \cite{4} extended their
isotropic solutions to anisotropic one for static spherically symmetric source and presented the graphical behavior of anisotropic factor.
Ovalle \cite{5} employed gravitational decoupling Via MGD and extracted anisotropic solutions from perfect fluid spheres.
The impact of charge on anisotropic
spherically symmetric solutions was measured through this approach assuming Krori-Barua solution \cite{6} and extended
their work to cylindrical symmetry \cite{7}. They have also examined the stability as
well as energy conditions in order to confirm the validity of their solutions.
Adopting the same approach, Gabbanelli et al. \cite{8} developed some new
anisotropic solutions using Durgapal-Fuloria stellar system which meet the criterion of physical admissibility.
Graterol \cite{9} considered Buchdahl perfect fluid distribution as an interior of a star and worked out
 anisotropic exact spherical solutions using MGD technique.

In \cite{10}, Durgapal's fifth isotropic solution which describes static case of
spherically symmetric fluid distribution has been considered and extended
to an anisotropic scenario with the help of MGD approach. Keeping in view the significance of this
approach, Estrada and Prado \cite{11} provided the higher-dimensional extension of gravitational decoupling method.
In \cite{12, 13}, Sharif and Waseem devoted their study to obtain spherically symmetric anisotropic and charged anisotropic
solution in modified theory of gravity. Using the gravitational decoupling through MGD approach,
an anisotropic version of Tolman VII solution was built and
an exact two-fluid solution for interior of stellar object was determined which is physically acceptable
and can predicts the behavior of compact objects \cite{14}.
An algorithm has been provided in \cite{14*} to show the decoupling of a gravitational source in pure Lovelock gravity.
MGD approach has been adopted to discuss different issues for 2+1 circularly symmetric and static spacetimes \cite{14**},
while an extended version of MGD method has been introduced in \cite{14***}. Gabbanelli et al. \cite{p} considered simple but generic isotropic Durgapal–Fuloria stars and extended them to the anisotropic domain via MGD approach, while in \cite{q}, the Maurya-Gupta isotropic fluid solution has been extended to an anisotropic domain. In cite{r, s}, a new deformed embedding class one solution has been explored for the realistic stars considering anisotropic and charged anisotropic fluid distributions, respectively.
Recently, Maurya et al. \cite{t} devoted their efforts to explore the possibility of getting solutions for ultradense anisotropic stellar systems with the help of gravitational decoupling via MGD approach within the framework of $f(R, T)$ theory of gravity, while decoupling gravitational sources by MGD approach in the framework of Rastall gravity have been presented in \cite{u}. Sharif and Ama-Tul-Mughani employed extended version of geometric deformation decoupling method and developed the solution for anisotropic static sphere.

Anisotropic stellar configurations have unequal pressure distribution in radial and tangential directions,
which is mainly followed by the presence of mixture of different kinds of
fluids, different kinds of transformations \cite{1*}, magnetic effects, viscosity, rotation and super-fluids \cite{2*}.
Anisotropic solutions basically provide a realistic description
of relativistic objects under a variety of circumstances.
The presence of pressure anisotropy in the matter configuration
introduces several interesting features as if one has positive anisotropy factor (i.e., $P_t-P_r>0$),
then stellar configuration experiences a repulsive force which counterbalances
the effects of gravity. Thus, it facilitates the construction
of more compact objects as compared to isotropic fluid distribution \cite{3*,5*}.
Lemaitre was the first person who described pressure anisotropy in a
perfect fluid distribution \cite{6*}, while
 Bower and Liang \cite{7*} worked on locally anisotropic equations of state
 and manifests their significance. Ruderman \cite{8*} suggested the presence of pressure anisotropy in a very high density regions
 of nuclear matter, i.e., $\rho>10^17 kg/m^3$. The issues that appear during the explanation of a viable cosmology has been discussed in \cite{9*}. 

The decoupling of gravitational sources by MGD is not just an interesting technique,
yet in addition it possesses various attractive features that make it especially
appealing in the search of new spherically symmetric solutions for Einstein's field equations.
If we briefly review the methodology, we find that it starts with a usual spherically symmetric source
, which is combined with a more complex gravitational source. The process of combining the
usual source with some new gravitational sources can be repeated multiple times, rather it continues until system preserves its symmetry
and additional sources do not exchange the energy momentum tensor. We can summarize it mathematically as

 \begin{eqnarray}\label{1}
    &&\hat{T}_{\mu\nu}\mapsto \tilde{T}_{\mu\nu}^{1}= \hat{T}_{\mu\nu}+\gamma^{(1)}{T^{1}_{\mu\nu}},\\\label{2}
   && \tilde{T}^{(1)}_{\mu\nu}\mapsto \tilde{T}_{\mu\nu}^{(2)}= \tilde{T}^{(1)}_{\mu\nu}+\gamma^{(2)}{T^{2}_{\mu\nu}},\\\nonumber
    &&.\\\nonumber
    &&.\\\nonumber
    &&.\\\label{2}
    &&\tilde{T}^{(n-1)}_{\mu\nu}\mapsto \tilde{T}_{\mu\nu}^{(n)}= \tilde{T}^{(n-1)}_{\mu\nu}+\gamma^{(n)}{T^{n}_{\mu\nu}},
 \end{eqnarray}
where $\hat{T}_{\alpha\beta}$ represents energy momentum tensor for simple gravitational source and $\tilde{T}_{\alpha\beta}$ represents
combination of energy momentum tensors for simple gravitational source and some other complex gravitational source $T_{\alpha\beta}$. It works until we have
\begin{eqnarray}\label{3}
    \nabla_{\beta} T^{\mu\nu} = \nabla_{\nu}T^{(1) \mu\nu }=.....= \nabla_{\nu}T^{(n) \mu\nu }=0.
\end{eqnarray}
For the solution of decoupled field equations, Einstein's field equations are solved
 for each source term separately and then all the solutions are combined in order to get the solution related to the
 total energy momentum tensor.

Among perfect fluid solutions presented by Tolman \cite{Tol}, we have chosen Tolman V solution which represents the fluid
spheres with infinite pressure and energy density at the center. In \cite{4, 14}, MGD approach was used to establish anisotropic
version of Tolman IV and VII which represent the fluid spheres with some definite values of pressure and density at the center.
The different behavior Tolman V solution regarding pressure and density at the center motivates us to
extend the Tolman V perfect fluid solution to anisotropic one via MGD approach and to analyze viability of new anisotropic solution. Undoubtedly, such type of solutions are not feasible for the representation of realistic stars, however the stars which are vulnerable to gravitational collapse can successfully be analyzed through such type of solutions. Gravitational collapse is a significant cosmological phenomenon in the sense where it is responsible for the end of a star, at the same time it gives birth to new stars.

This paper has been arranged as follows: In the next section, we present decoupling
field equations. In sections III and IV we discuss MGD technique and their
interrelated junction conditions. Section V presents discussion about the Tolman V perfect fluid solution
and some useful results obtained by applying matching conditions. Section VI explores new anisotropic solution
which is followed by the physical analysis of the solution. Last section summarizes the results.

\section{Gravitational Decoupling of Einstein's Field Equations}

In the the framework of decoupling method developed by Ovalle \cite{5}, the Einstein's field equations take
the form as
\begin{eqnarray}\label{ei}
  R_{\mu\nu}-\frac{1}{2}Rg_{\mu\nu}&=&-\kappa^2 T_{\mu\nu}^{(total)},
\end{eqnarray}
where $T_{\mu\nu}^{(total)}$ represents the sum of energy momentum tensor (EMT) for perfect fluid and for some other source, i.e.,
\begin{eqnarray}\label{tot}
  T_{\mu\nu}^{(total)} &=& T_{\mu\nu}^{(perfect fluid)}+T_{\mu\nu}^{(other source)}.
\end{eqnarray}

EMT for perfect fluid has the following expression
\begin{eqnarray}\label{5}
T_{\mu\nu}^{(perfect fluid)}&=&({\rho}+P){U_{\mu}}{U_{\nu}}-Pg_{{\mu}{\nu}},
\end{eqnarray}

 where ${\rho}$ and $P$ represent energy density and isotropic pressure of the fluid respectively,
 while $U^{\mu}=e^{-\frac{\xi}{2}}\delta_0^\mu$ is the four velocity of fluid which satisfies the relation ${U_{\mu}}{U^{\mu}}=1$.

We consider an additional source $\phi_{\mu\nu}$ in EMT by using MGD approach which may be a scalar,
vector or tensor field but responsible for the anisotropy in the fluid. Under this consideration, Eq.$(\ref{tot})$ becomes
\begin{eqnarray}\label{7}
T^{total}_{\mu\nu}&=&T^{pf}_{\mu\nu}+\beta \phi_{\mu\nu},
\end{eqnarray}
where, $\beta$ is an intensity parameter. After this modification,
Eq.(\ref{ei}) can be rewritten as
\begin{eqnarray}\label{8}
G_{\mu\nu}&=&T^{perfect fluid}_{\mu\nu}+\beta\phi_{\mu\nu},
\end{eqnarray}
with Einstein tensor $G_{\mu\nu}=R_{\mu\nu}-\frac{1}{2} R g_{\mu\nu}$ .

 We have considered spherically symmetric static gravitational source whose interior geometry is represented by
\begin{eqnarray}\label{10}
    ds^{2}&=& e^{\xi(r)}dt^{2}-e^{\eta(r)}dr^{2}-r^2(d\theta^{2}+\sin^{2}\theta d\phi^{2}),
\end{eqnarray}
where, $\xi(r)$ and $\eta(r)$ are the functions of radial coordinate ranging $0<r<R$.
The Einstein's filed equations for the case under consideration take the form as
\begin{eqnarray}\label{11}
\kappa^2\left(\rho+\beta \phi^{0}_{0}\right)&=&e^{-\eta}\Big(\frac{\eta'}{r}-\frac{1}{r^2}\Big)+\frac{1}{r^2},\\\label{12}
\kappa^2\left(P-\beta \phi^{1}_{1}\right)&=&e^{-\eta}\Big(\frac{\xi'}{r}+\frac{1}{r^2}\Big)-\frac{1}{r^2},\\\label{13}
\kappa^2\left(P-\beta \phi^{2}_{2}\right)&=&e^{-\eta}\Big(\frac{\xi''}{2}-\frac{\eta'}{2r}+\frac{\xi'}{2r}-\frac{\xi'\eta'}{4}+\frac{\xi'^2}{4}\Big).
\end{eqnarray}

As MGD approach does not affect the validity of conservation law, so
the covariant divergence of EMT given in Eq.$(\ref{7})$ gives the following output
\begin{eqnarray}\label{14}
P'+ \frac{{\eta}'}{2}(\rho+P)
-\beta{(\phi)^{1}_{1}}'
+\frac{{\xi}'\beta}{2}(\phi^{0}_{0}-\phi^{1}_{1})+\frac{2\beta}{r}(\phi^{2}_{2}-{\phi}^{1}_{1})=0.
 \end{eqnarray}

Here, prime stands for the derivative with respect to radial coordinate.
The set of equations (\ref{11})-(\ref{13}) comprise of seven unknowns, namely
$\rho$, $P$ (physical variables), $\xi$, $\eta$  (geometric functions), and three
independent components appearing due to the additional source $\phi_{\mu\nu}$. The solution
of field equations required the evaluation of aforementioned unknown functions which can be done
 following the analytical approach introduced by Ovalle. Before proceeding towards the implementation
MGD technique, it is worthwhile to define our physical parameters
i.e., $\tilde{\rho}$, $\tilde{P_{r}}$ and $\tilde{P_{t}}$ as
\begin{eqnarray}\label{28}
\tilde{\rho}&=&\rho+\beta \phi^{0}_{0},\\\label{29}
\tilde{P_{r}}&=&P-\beta\phi^{1}_{1},\\\label{30}
\tilde{{P}_{t}}&=&P-\beta\phi^{1}_{1}.
\end{eqnarray}
To find the unknowns present in Eqs.(\ref{11})-(\ref{13}), we follow the analytical
approach introduced by Ovalle \cite{5}. It can be seen clearly that the source $\phi_{\mu\nu}$ gives rise to the emergence of
anisotropy in the interior of a self gravitational system, thus anisotropic factor assumes the following expression
\begin{eqnarray}\label{31}
\tilde{\Delta}&=&\tilde{P_{t}}-\tilde{P_{r}}=\beta(\phi^{1}_{1}-\phi^{2}_{2}).
\end{eqnarray}

\section{Minimal geometric deformation}
Now, we follow MGD approach in order to decouple the Einstein's filed equations given in
(\ref{11})-(\ref{13}). In this methodology, the framework is changed in such a way that field equations
related with the source term $\phi_{\mu\nu}$ appear as quasi Einstein equations. For this, we assume
$\beta=0$ to find the perfect fluid solution $(\xi, \eta, \rho, P)$ in the framework of the line element given as
\begin{eqnarray}\label{35}
    ds^2&=&e^{\mu(r)}dt^{2}-\frac{dr^{2}}{\nu(r)}-r^{2}(d\theta^{2}+\sin^2\theta d\phi^{2}),
\end{eqnarray}
here $\nu(r)= 1- \frac{2m}{r}$ with $m$ as a mass function.
Now, the impact of intensity parameter $\beta$ on the additional source for the perfect
fluid variables $(\xi, \eta, \rho, P)$ encodes geometric decomposition experienced by perfect fluid geometry as
\begin{eqnarray}\label{36}
    \mu\mapsto \xi&=& \mu+\beta k,\\\label{37}
    \nu\mapsto e^{-\eta}&=&\nu+\beta h,
\end{eqnarray}
where $h$ and $k$ are the deformations experienced by the radial and temporal metric coefficients, respectively.
These deformations are responsible for the emergence of anisotropy in the fluid configuration. As we are following the
minimal geometric deformation which is based on the conditions
\begin{eqnarray}\label{38}
    k&\mapsto& 0,
\end{eqnarray}
and
\begin{eqnarray}\label{39}
    h&\mapsto &h^{*}.
\end{eqnarray}
Here, we introduce the minimal deformation only against the radial component
while the temporal one experiences no change. Thus, we have
\begin{eqnarray}\label{40}
\mu\mapsto&=&\mu , \nu\mapsto e^{-\eta}=\nu+\beta h^{*}.
\end{eqnarray}
Now, we consider the deformed matric and extract two set of equations from Eqs.(\ref{11})-(\ref{13}). First one is obtained against
 $\beta=0$ and comprises Einstein's field equations for perfect fluid, while the second one involves additional source $\phi_{\mu}{\nu}$.
 The first set of equations is given as
\begin{eqnarray}\label{41}
\kappa^2\rho&=&\frac{1}{r^2}-\Big(\frac{\nu}{r^2}+\frac{{\nu}'}{r}\Big),\\\label{42}
\kappa^2P &=&-\frac{1}{r^{2}}+\frac{\nu}{r}\Big(\frac{1}{r}+{\xi}'\Big),\\\label{43}
\kappa^2P&=&\frac{\nu}{4}\Big(2{\xi}''+{{\xi}'}^{2}+\frac{2{\xi}'}{r}\Big)+\frac{{\nu}'}{4}\Big(
    {{\xi}'}+\frac{2}{r}\Big),
\end{eqnarray}
while the second set of quasi-equations is given by
\begin{eqnarray}\label{44}
\kappa^2\phi^{0}_{0}&=&-\frac{{h^*}'}{r}
-\frac{h^{*}}{r^2},\\\label{45}
\kappa^2\phi^{1}_{1}&=&-\frac{h^{*}}{r}\Big(\frac{1}{r}+{\xi}'\Big),\\\label{46}
\kappa^2\phi^{2}_{2}&=&-\frac{h^{*}}{4}\Big(2{\xi}''+{{\xi}'}^{2}+\frac{2{\xi}'}{r}\Big)-\frac{h^{*'}}{4}
    \Big({\xi}'+\frac{2}{r}\Big).
\end{eqnarray}
However, the conservation equations for the both scenarios take the form as
\begin{eqnarray}\label{c1}
  P'+\frac{\eta'}{2}(\rho+P) &=& 0,\\\label{c2}
 ( \phi_1^1)'-\frac{\eta'}{2}(\phi_0^0-\phi_1^1)-\frac{2}{r}(\phi_2^2-\phi_1^1)&=&0.
\end{eqnarray}
It is worthwhile to mention here that both of Eqs.(\ref{c1}) and (\ref{c2}) have to obey the general conservation law.
More precisely, both systems do not have dependence on each other for conservation and their interaction is purely gravitational.
 Also,  Eqs.(\ref{44})-(\ref{46}) with additional
source $\phi_{\mu\nu}$ ($\phi^{0}_{0}= \rho$, $\phi^{1}_{1}=P_{r}$, $\phi^{2}_{2}=P_{t}$)
look similar to the field equations for the metric
\begin{eqnarray}\label{47}
    ds^{2}&=&e^{\mu(r)}dt^{2}-\frac{dr^{2}}{h^*(r)}-r^{2}(d\theta^{2}+\sin^{2}\theta).
\end{eqnarray}
However, the right hand sides of Eqs.(\ref{45}) and (\ref{46}) differ by $\frac{1}{r^{2}}$ for the anisotropic
solutions which can be read as
\begin{eqnarray}\label{48}
\tilde{\rho}&=&{\phi^{0}_{0}}^{*}=\phi^{0}_{0}+\frac{1}{r^{2}},\\\label{49}
\tilde{P_{r}}&=&{\phi^{1}_{1}}^{*}=\phi^{1}_{1}+\frac{1}{r^{2}},\\\label{50}
\tilde{P_{t}}&=&{\phi^{2}_{2}}^{*}=\phi^{2}_{2}={\phi^{3}_{3}}^{*}=\phi^{3}_{3},
\end{eqnarray}

\section{Junction Condition}

Junction conditions plays a vital role in the study of stellar geometry by matching the
 interior and exterior surfaces of a star. The smooth matching of the surfaces is significantly helpful in the investigation of
  stellar configuration.
The interior geometry for our case following the MGD approach is given by
\begin{eqnarray}\label{51}
    ds^{2}&=&e^{-\xi(r)_{+}}-\Big(1-\frac{2\tilde{m}(r)}{r}\Big)^{-1}dr^{2}- r^{2}(d\theta^{2}+\sin^{2}\theta d\phi^{2}),
\end{eqnarray}
where $\tilde{m}=m(r)-\frac{\beta r}{2}h^{*}(r)$ represent interior mass of the system and $h^*$ is given in Eq.(\ref{40})
and yet to be evaluated later. The metric for interior geometry should be matched with one for the
exterior geometry, where we assume that $\rho^+ =P^+=0$.
However, the exterior geometry is no more vacuum as it is influenced by the additional gravitational source $\phi_{\mu\nu}$.
The general metric for exterior geometry can be written as
\begin{eqnarray}\label{52}
    ds^{2}&=& e^{\xi(r)_{+}}-e^{\eta(r)_{+}}dr^{2}- r^{2}(d\theta^{2}+\sin^{2}\theta d\phi^{2}),
\end{eqnarray}
where  $\xi(r)$ and $\eta(r)$ can found by solving
\begin{eqnarray}\label{s}
  R_{\mu\nu}-\frac{1}{2}Rg_{\mu\nu} &=& -\kappa^2 \beta\phi_{\mu\nu}.
\end{eqnarray}
The continuity of first fundamental form on the boundary of star $r=R$ leads towards the expressions
\begin{eqnarray}\label{53}
    \xi(r)^{+}&=&\xi(r)^{-}~~  and ~~ 1-\frac{2M_{0}}{R}+ \beta h^{*}(R)= e^{-\eta(R)^{+}},
\end{eqnarray}
which follow from the condition $[ds^{2}]_{\sum}=0$ with $V_{\sum}=V(R)^{+}-V(R)^{-}$, where $V=V(R)$ be any arbitrary function.
$h^{*}(R)$ and $M_{0}=m(R)$ represent the total deformation and mass at the star's surface respectively.
From the continuity of second fundamental form, we have
\begin{eqnarray}\label{g}
 [G_{\mu\nu}r^\mu]_\Sigma &=& 0,
\end{eqnarray}
 $r^{\mu}$ represents unit four vector in radial direction. The expression given in Eq.(\ref{ei}) together with Eq.(\ref{g})
 yields
 \begin{eqnarray}\label{t}
 [T_{\mu\nu}^{(total)}r^\mu]_\Sigma  &=& 0,
\end{eqnarray}
which further provides
\begin{eqnarray}\label{54}
P_R-\beta(\phi^{1}_{1}(R))^{-}&=&-\beta(\phi^{1}_{1}(R))^{+}.
\end{eqnarray}
Here, $P_R=P^-(R)$. With the help of Eq.(\ref{45}), we reach at
\begin{eqnarray}\label{55}
P_{R}+ \frac{\beta h^{*}(R)}{\kappa^2R}\Big({\xi}'+\frac{1}{R}\Big)&=&\frac{\beta k^{*}(R)}{\kappa^2R}\Big(\frac{1}{R}+\frac{2M}{R(R-2M)}\Big),
\end{eqnarray}
where choice of $k^{*}(R)$ is radial geometric deformation for Schwarzschild metric under the influence of
additional source term $\phi_{\mu\nu}$ which provides
\begin{eqnarray}\label{56}
    ds^{2}&=&\Big(1-\frac{2M}{r}\Big)- \Big(1-\frac{2M}{r}+\beta k^{*}\Big)^{-1} dr^{2}-  r^{2}(d\theta^{2}+\sin^{2}\theta d\phi^{2}).
\end{eqnarray}
The necessary and sufficient condition for the matching of interior MGD metric and the deformed Schwarzschild
metric are given in the Eqs.(\ref{52}) and (\ref{56}). Now, we consider the the exterior metric as standard "Schwarzschild metric"
which requires $k^{*}=0$ in Eq.(\ref{55}) and provides
\begin{eqnarray}\label{57}
 \tilde{P}_{R}&=& P_{R}+ \frac{\beta h^{*}(R)}{R}\Big(\xi'+\frac{1}{R}\Big) =0.
\end{eqnarray}
Here, we have a significant result that anisotropic radial pressure vanishes at the surface which is required for
the equilibrium of a star in original Schwarzschild solution.

\section{Interior perfect fluid for Tolman V solution}
Now, we describe the properties of Tolman V perfect fluid solution \cite{Tol} for which we intend to develop anisotropic
solution through MGD approach. Tolman V solution is basically natural one for the description of
the spheres of fluid with infinite pressure and density at the center. We choose a solution whose
physical characteristics are different from Tolman IV and VII.
 Its metric coefficients are
given by
\begin{eqnarray}\label{58}
    e^{\xi(r)}&=&B^{2}r^{2n}, \\\label{59}
    e^{\eta(r)}&=&\nu^{-1} =\frac{1+2n-n^{2}}{1-(1+2n-n^{2})\Big(\frac{r}{C}\Big)^{N}},
 \end{eqnarray}
where $N =\frac{2(1+2n-n^{2})}{1+n}$. The constants appearing in Eqs.(\ref{58}) and (\ref{59}) can be
obtained by matching conditions between the interior and exterior solutions and are given in Eqs.(\ref{62})-(\ref{64}).
Inside the system, density $\rho$ and pressure $P$ are given by the expressions
\begin{eqnarray}\label{60}
    \rho&=&\frac{2n-n^{2}}{8\pi r^{2}(1+2n-n^{2})}+\frac{3+5n-2n^{2}}{8\pi(1+n)C^{2}}\Big(\frac{r}{C}\Big)^L,\\\label{61}
    P&=&\frac{n^{2}}{8\pi r^{2}(1+2n-n^{2})}-\frac{1+2n}{8\pi C^{2}}\Big(\frac{r}{C}\Big)^L,
\end{eqnarray}
where $L = \frac{2n(1-n)}{1+n}$.
Here, an analytical relationship is possible to find that connects the density and pressure inside the fluid sphere. Thus we have the
equation of state given below
\begin{eqnarray}\label{equ}
 \rho &=&  \frac{2n-n^2}{n^2}P+\frac{1+4n+3n^2-2n^3}{4\pi n(n+1)C^N}\left(\frac{n}{4\pi(\rho+3P+n\rho-nP)}\right)^{\frac{L}{2}}.
\end{eqnarray}
At the center, density and pressure become infinite but their ratio $\frac{pr}{\rho}$ satisfy the Zeldovichi's condition (i.e., $\frac{pr}{\rho}<1$)
and attains the value $\frac{1}{3}$
A general expression for the boundary of the sphere where pressure is dropped to zero value takes the form as
\begin{eqnarray}\label{rad}
 r_b  &=& C\left(\frac{n^2}{(1+2n-n^2)(1+2n)}\right)^{\frac{1}{N}}.
\end{eqnarray}
At the value of $r_b$ where pressure is dropped to zero, the density of fluid configuration takes the form as
\begin{eqnarray}\label{den}
  \rho_b &=&  \frac{1}{8\pi C^2}\left(\frac{2n+4n^2-2n^3}{(1+2n-n^2)(1+n)}\right)\left(\frac{n^2}{(1+2n-n^2)(1+2n)}\right)^{-\frac{2}{N}}.
\end{eqnarray}
 Using the matching conditions given in previous section, we have worked out the values of constants appearing in above equations as
 \begin{eqnarray}\label{62}
    B^{2}&=&\frac{M_{0}}{nR^{2n+1}},\\\label{63}
    n&=&\frac{M_{0}}{R-2M_{0}},\\\label{64}
  \Big(\frac{R}{C}\Big)^{N}&=&\frac{M_{0}(M_{0}R-2M_{0}^{2})}{R_0(R^{2}-2M_{0}R-M_{0}^{2})},
 \end{eqnarray}
with $\frac{M_{0}}{R}< \frac{4}{9}$. Here, $M_0=m(R)$ represents total mass of fluid distribution. Further, the expression developed for $n$
imposes the condition that $R\neq 2M_0$ otherwise it leads to $n\rightarrow\infty$.
The above expressions (\ref{62})-(\ref{64}) ensure that the continuity of geometry will not change
but in the presence of additional source $\phi_{\mu\nu}$ at the boundary $r=R$ of star.

Next, we explore the anisotropic solutions by gravitational decoupling for some non-zero value of $\beta$.
 The explicit expression of
metric components are given by Eqs.(\ref{40}) and (\ref{58}) , whereas interior deformation $h^{*}(r)$ and the
source term $\phi_{\mu\nu}$ are connected through Eqs.(\ref{44})-(\ref{46}).

\section{Anisotropic Solution for Tolman V}

One can see that the exterior geometry of Schwarzschild metric is compatible with interior configuration
if $\beta{\phi^{1}_{1}}(R) \sim P(R)$. Following this mimic constraint, we can make the choice as
\begin{eqnarray}\label{65}
   \phi^{1}_{1}&=&P ~\Rightarrow~ h^{*}=-\nu(r)+\frac{1}{r\xi'+1},
\end{eqnarray}
which can further be written using using Eq.(\ref{45}) as
\begin{eqnarray}\label{66}
 h^{*}&=&-\nu(r)+\frac{1}{r\xi'+1}.
\end{eqnarray}
It helps us to find the radial component as follows
\begin{eqnarray}\label{67}
    e^{-\eta(r)}&=&\nu(r)(1-\beta)+\frac{\beta}{2n+1}.
\end{eqnarray}
The Eqs.(\ref{58}) and (\ref{67}) give the Tolman V solution which is
minimally deformed by generic anisotropic source $\phi_{\mu\nu}$. For $\beta=0$, Eq.(\ref{67}) is turned into the form that represents
original Tolman V solution \cite{5}. Our next task is to match the interior and exterior geometry for new solution.
The continuity of first fundamental form generates the following result
\begin{eqnarray}\label{68}
    B^{2}R^{2n}&=&1-\frac{2M}{R},\\\label{69}
     \nu(r)(1-\beta)+\frac{\beta}{2n+1}&=&1-\frac{2M}{R},
\end{eqnarray}
whereas the continuity of second fundamental form $P(R)=0$ using Eq.(\ref{61}) leads to
\begin{equation}\label{70}
    \Big(\frac{n^{2}}{1+2n-n^{2}}\Big)=\frac{(1+2n)R^{2}}{C^2}\Big(\frac{R}{C}\Big)^{L}.
\end{equation}
Now, using Eq.(\ref{53}) together with Eq.(\ref{69}), we obtain Schwarzschild Mass
\begin{equation}\label{71}
    \frac{2M}{R}=\frac{2M_{0}}{R}-\beta\Big(1-\frac{2M_{0}}{R}+\frac{1}{2n+1}\Big),
\end{equation}
which further implies that
\begin{equation}\label{72}
      B^{2}R^{2n}= 1- \frac{2M_{0}}{R}+\beta\Big(1-\frac{2M_{0}}{R}+\frac{1}{2n+1}\Big).
\end{equation}
 Eqs.(\ref{68})-(\ref{72}) provide necessary and sufficient conditions for the smooth matching of new anisotropic inner solution
 and outer Schwarzschild solution at the
 boundary of the star. Using the constraint given by Eq.(\ref{65}) in Eqs.(\ref{28})-(\ref{30}), we are able to develop the expressions for physical
parameters ($\tilde{\rho}$, $\tilde{P_{r}}$ and $\tilde{P_{t}}$) as
\begin{eqnarray}\label{73}
\tilde{\rho}&=&\frac{(2-n)n}{8\pi(1+2n-n^2)r^{2}}-\frac{(n-3)(1+2n)(\frac{r}{C})^{L}}{8\pi
   C^{2}(1+n)}+\frac{n^2-(1+2n)(1+2n-n^2)(1+N)(\frac{r}{C})^{N})}{8\pi(1+2n)(1+2n-n^2)r^2}\beta,
\\\label{74}
\tilde{P_{r}}&=&(1-\beta)\Big(\frac{n^{2}}{8\pi r^{2}(1+2n-n^2)}-\frac{(1+2n)}{8\pi C^{2}}
(\frac{r}{C})^{L}\Big),
\\\nonumber
\tilde{P_{t}}&=&\frac{n^{2}}{8\pi(1 +2n-n^2)r^{2}}-\frac{(1+2n)(\frac{r}{C})^{L}}{8\pi C^{2}}
-\beta\frac{n^2}{r^2}\left(\frac{n^2-(1+2n)(1+2n-n^2)(\frac{r}{C})^N}
{8\pi(1+2n-n^2)(1+2n)}\right)\\\label{75}&&+\frac{\beta}{16\pi r^2}N(n+1)(\frac{r}{C})^N,
\end{eqnarray}

\begin{figure}
\centering \epsfig{file=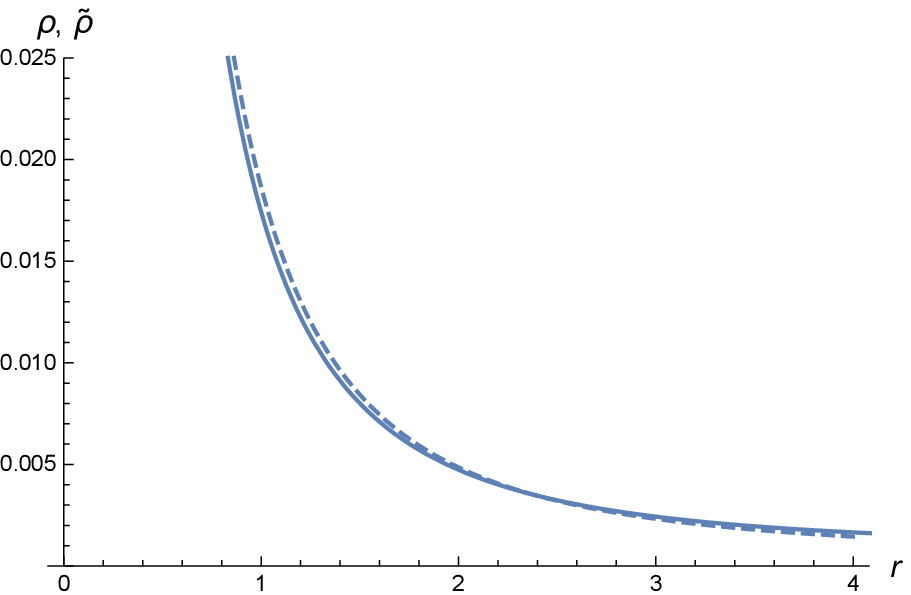, width=.4\linewidth,
height=1.5in}\epsfig{file=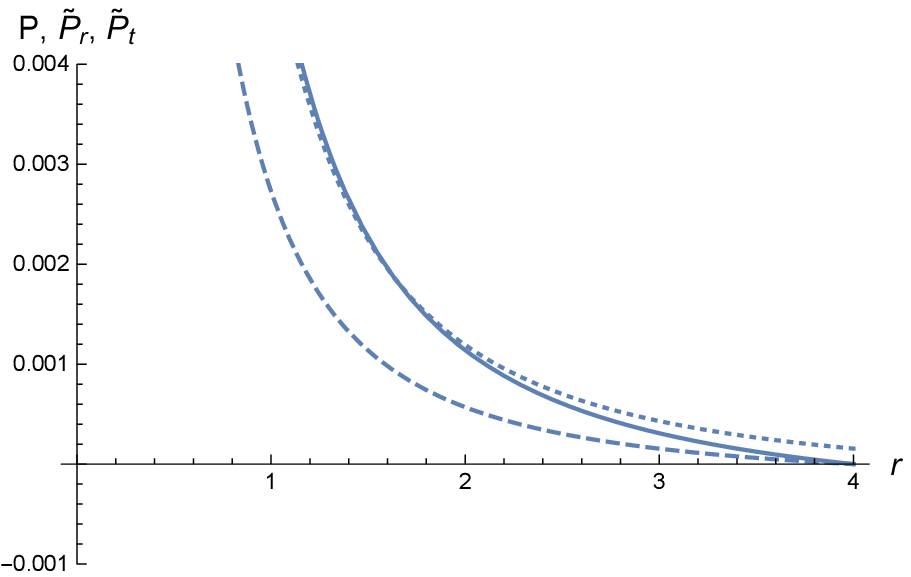, width=.4\linewidth,
height=1.5in} \caption{\label{Fig.1} shows the graphs of physical quantities for the new anisotropic solution and Tolman V perfect fluid solution
. Herein, we set $R=4$ and $M_0=1$ with $\beta=0.5$. In left panel curves represent $\rho$ (Solid), $\tilde{\rho}$ (Dashed) whereas right panel $P$ (solid), $\tilde{P_r}$ (Dashed) and $\tilde{P_t}$ (Dotted).}
\end{figure}

Left sided penal of the Fig.\ref{Fig.1} shows the comparison of energy density profile for the perfect fluid Tolman V solution and new anisotropic
solution, while the right sided penal of Fig.\ref{Fig.1} exhibits the comparison of pressure profile for the perfect fluid Tolman V and new anisotropic
solution. We have fixed the values $R=4$ and $M_0=1$. We see from the figures that all the physical quantities
decrease monotonically. We can also see that energy density attains finite value at the boundary while it becomes infinite at the center. As a
physically valid stellar model always obeys the condition that radial component of pressure must vanish at the surface which can clearly be observed
from Fig.\ref{Fig.1}. It is also noteworthy that tangential pressure gets positive value throughout in stellar configuration.

\subsection{Anisotropy Measurement and Equation of States Parameters}

\begin{figure}
\centering \epsfig{file=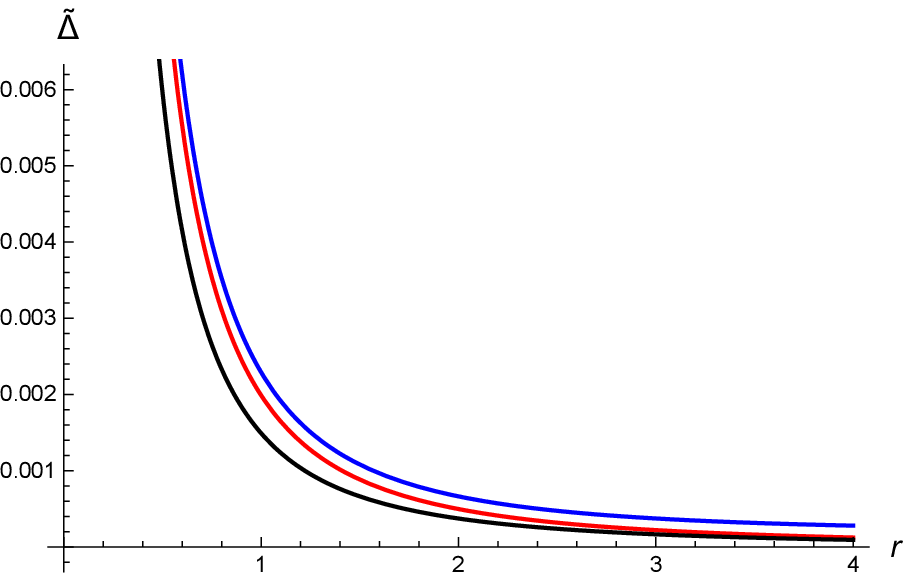, width=.4\linewidth,
height=1.5in}\epsfig{file=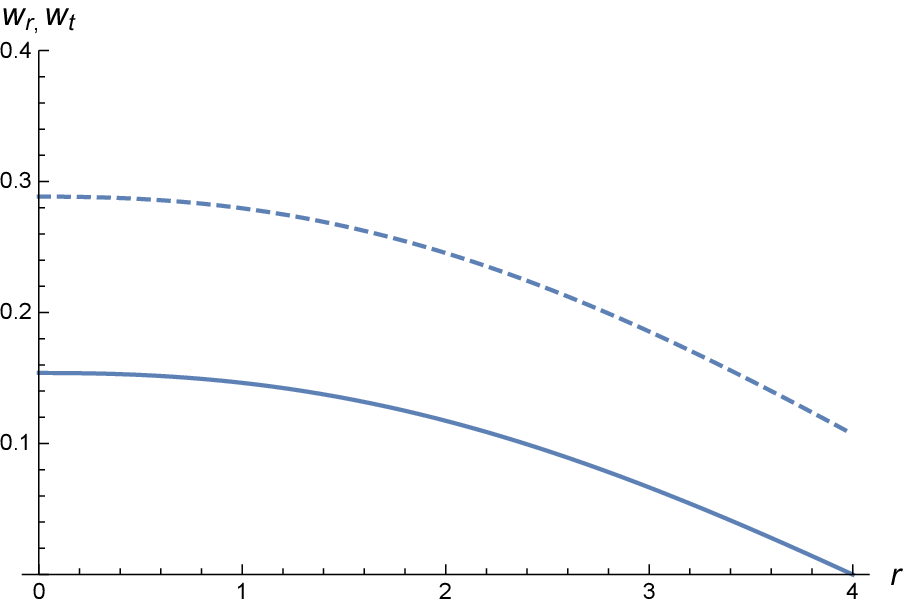, width=.4\linewidth,
height=1.5in} \caption{\label{Fig.2} Left penal presents dependence of anisotropy against different values of $\beta$, i.e.,
$\beta=0.5$ (Blue), $\beta=0.4$ (Red), $\beta=0.3$ (Black).
Right penal comprises variation of EoS parameters for radial and tangential components, $w_r$ (Solid), $w_t$ (dashed).}
\end{figure}

Here, anisotropic factor $\tilde{\Delta}$ defined in Eq.(\ref{31}) takes the form as
\begin{eqnarray}\nonumber
\tilde{\Delta}&=&-\beta\frac{n^2}{8\pi r^2}\left(\frac{n^2-(1+2n)(1+2n-n^2)(\frac{r}{C})^N}
{(1+2n-n^2)(1+2n)r^2}\right)+\frac{\beta}{16\pi r^2}N(n+1)(\frac{r}{C})^N\\\label{76}&&
+\Big(\frac{n^{2}}
{8\pi(1+2n-n^2)}-\frac{(1+2n)}{8\pi C^{2}}(\frac{r}{C})^{L}\Big)\beta.
\end{eqnarray}
In order to measure the anisotropy effects for our new Tolman V solution, we have drawn the graphs for different values $\beta$
in left penal of Fig.\ref{Fig.2}, where we can observe that anisotropy factor $\tilde{\Delta}$ and intensity factor $\beta$ have direct relation. As the value
of $\beta$ is increased, anisotropy is also increased. We also obtain $\tilde{\Delta}>0$ for all values of $\beta$ which depicts $\tilde{P_{t}}>\tilde{P_{r}}$.
It facilitates the construction of more compact objects. However, the choice of negative value for $\beta$ will lead to the result $\tilde{\Delta}<0$.

However, EOS parameters for radial and tangential components, i.e., $w_r=\frac{\tilde{P}_r}{\tilde{\rho}}$ and $w_t=\frac{\tilde{P}_t}{\tilde{\rho}}$ respectively,
assume the following form
\begin{eqnarray}\label{73*}
w_r&=& \frac{(1-\beta)\Big(\frac{n^{2}}{r^{2}(1+2n-n^2)}-\frac{(1+2n)}{ C^{2}}
(\frac{r}{C})^{L}\Big)}{\frac{(2-n)n}{(1+2n-n^2)r^{2}}-\frac{(n-3)(1+2n)(\frac{r}{C})^{L}}{
   C^{2}(1+n)}+\frac{n^2-(1+2n)(1+2n-n^2)(1+N)(\frac{r}{C})^{N})}{(1+2n)(1+2n-n^2)r^2}\beta},
\\\nonumber
w_t&=&\frac{\frac{n^{2}}{8\pi(1 +2n-n^2)r^{2}}-\frac{(1+2n)(\frac{r}{C})^{L}}{8\pi C^{2}}
-\beta\frac{n^2}{r^2}\left(\frac{n^2-(1+2n)(1+2n-n^2)(\frac{r}{C})^N}
{8\pi(1+2n-n^2)(1+2n)}\right)+\frac{\beta}{16\pi r^2}N(n+1)(\frac{r}{C})^N}{\frac{(2-n)n}{(1+2n-n^2)r^{2}}-\frac{(n-3)(1+2n)(\frac{r}{C})^{L}}{
   C^{2}(1+n)}+\frac{n^2-(1+2n)(1+2n-n^2)(1+N)(\frac{r}{C})^{N})}{(1+2n)(1+2n-n^2)r^2}\beta}.\\\label{74*}
\end{eqnarray}

The graphical behavior of EOS parameters is evident in right panel of Fig.\ref{Fig.3}. It can be seen that both of the
parameters attain their maximum values at the center and minimum value at the surface of the sphere fluid.

\section{Physical Analysis}

In this section, we discuss some physical properties of the new anisotropic solutions in order to check their physical viability.

\subsection{Energy Conditions}

Energy conditions are often required to check the physical acceptability of different
significant results of cosmological geometries and gravitational fields, for example the ``no hair theorem" or the ``laws of BH thermodynamics".
Matt Visser and Carlos Barcello \cite{En} discussed cosmological implications of energy conditions
and they found that there are some quantum and relativistic effects that may violate all energy conditions
that comes up with new possibilities and enhances the significance of energy conditions.
Here, we check if these conditions are satisfied for our new anisotropic solutions.
Four explicit forms of energy conditions are given by\\
NEC: $\tilde{\rho}+\tilde{P_{t}}\geq 0, \tilde{\rho}+\tilde{P_{r}}\geq 0$.\\
WEC: $\tilde{\rho}\geq 0, \tilde{\rho}+\tilde{P_{t}}\geq 0, \tilde{\rho}+\tilde{P_{r}}\geq 0$.\\
SEC: $\tilde{\rho}+\tilde{P_{t}}\geq 0, \tilde{\rho}+\tilde{P_{r}}\geq 0, \tilde{\rho}+\tilde{P_{r}}+2\tilde{P_{t}}\geq 0$.\\
DEC: $\tilde{\rho}-\mid\tilde{P_{t}}\mid\geq 0, \tilde{\rho}-\mid\tilde{P_{r}}\mid\geq 0$.\\

\begin{figure}
\centering \epsfig{file=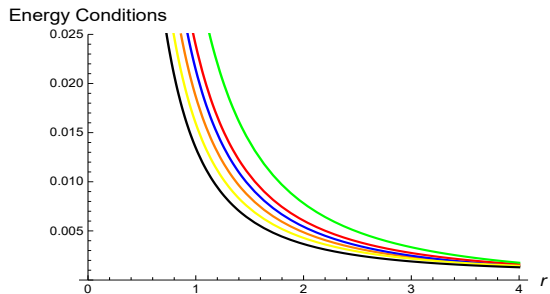, width=.4\linewidth,
height=1.5in} \caption{\label{Fig.3} presents variations of energy conditions for $\beta=0.5$, i.e., NEC (Blue and Red), WEC (Orange), SEC (Green), DEC (Yellow and Black).}
\end{figure}
Fig.\ref{Fig.3} exhibits the behavior of energy conditions that is found in agreement for anisotropic Tolman V solution.

\subsection{Casuality conditions}

\begin{figure}
\centering \epsfig{file=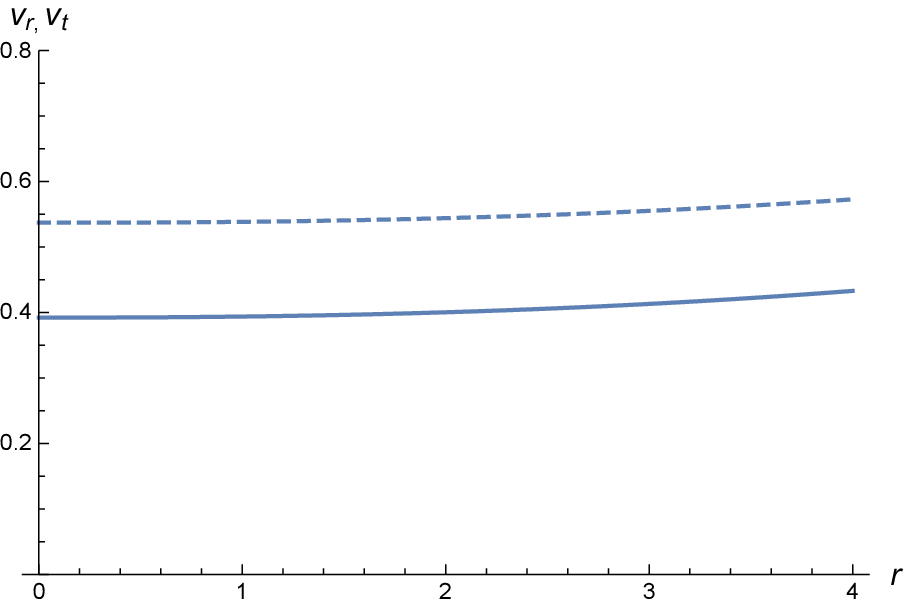, width=.4\linewidth,
height=1.5in}\epsfig{file=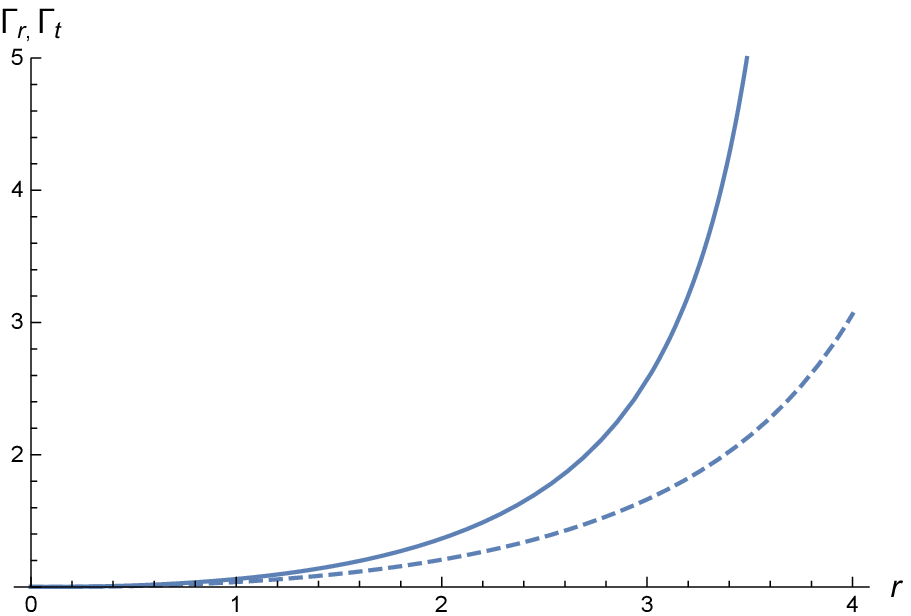, width=.4\linewidth,
height=1.5in} \caption{\label{Fig.4} Variation of casuality (left penal) and stability conditions (right penal), $v_r, \Gamma_r$ (Solid), $v_t, \Gamma_t$ (dashed).}
\end{figure}

One can manage the stability analysis of compact objects considering speed of sound \cite{s1}-\cite{s3}.
A physically accepted solution always obeys the rule that speed of light $c$ exceeds the
speed of sound $c_{s}$. This leads to rejecting the ``Low-energy effective field
theories" against to having ``Lorentz-invariant Lagrangian" by admitting the superluminal
variations \cite{s4}. Thus, the radial and transverse components of speed of sound, denoted
by $v_{r}$ and $v_{t}$, respectively, should be less than speed of light, which yields the inequalities

\begin{eqnarray}\label{inq}
0\leq v_{r}\leq 1,\quad\quad 0\leq v_{t}\leq 1,
\end{eqnarray}

where $v^2_{r}=\frac{dp_r}{d\rho}$, $v^2_{t}=\frac{dp_t}{d\rho}$ and $c=1$. Fig.\ref{Fig.4} (left penal) clearly shows that
speed of sound in both radial and tangential directions obeys the casuality condition which ensures the viability of the solution.

\subsection{Stability conditions}

Adiabetic index is a powerful tool when stability of stellar configuration is discussed.
For a spherically symmetric stable stellar configuration, it should possess the value greater than $\frac{4}{3}$.
The mathematical expressions of adiabetic index in radial and tangential directions takes the form as
\begin{eqnarray}\label{ss}
  \Gamma_r &=& \frac{d(\log P_r)}{d(log \rho)},\\\label{ss1}
   \Gamma_t &=& \frac{d(\log P_t)}{d(log \rho)}.
\end{eqnarray}

Fig.\ref{Fig.4} (right penal) shows that stability condition is not always being satisfied by both of the adiabetic indices.
It can clearly be seen that adiabetic index does not meet the stability criteria as it moves towards the center of the system,
which indicates that system under consideration may expose to gravitational collapse in future. This characteristic of the new solution
is inherited by its perfect fluid solution, however it becomes more crucial if anisotropy measure is increased.

\section{Conclusion}

MGD decoupling approach has been presented to incorporate the anisotropic effects of fluid distribution. In this paper,
 we have chosen well known Tolman V perfect fluid solution which offers exact solution for strange stars having infinite values
 of energy density and pressure at the center. We have employed MGD approach and developed new anisotropic solution. For this,
 we have decoupled Einstein's field equations in two sectors which has only gravitational interaction, one
 of them corresponds to the perfect fluid source, while the other
 is related to the anisotropic source.

 We have used mimic constraint for radial pressure, i.e., $\beta{\phi^{1}_{1}}(R) \sim P(R)$,
 which is based on matching condition given in Eq.(\ref{57}). Then, Tolman V perfect fluid solution
has been deformed by the general anisotropic source $\phi_{\mu\nu}$. The matching conditions
are worked out for the smooth matching of interior metric given in Eq.(\ref{10}) with metric coefficients presented in
sections V and VI and Schwarzchild outer space-time at the boundary of the sphere. The effects of anisotropy
has also been studied on Schwarzchild mass and found negligible when Schwarzchild mass `$M$' for anisotropic
version is compared with the mass of perfect fluid configuration `$M_0$'. The expressions for physical quantities
like energy density, radial and tangential pressure has been given in Eqs.(\ref{73})-(\ref{75}), whose graphical behavior
shows that anisotropic version of Tolman V solution is in agreement with the physical
properties of its parent solution. The mathematical expressions for anisotropy measurement and equation of state
have been explored. The graphical behavior of anisotropy in Fig.\ref{Fig.2} shows its dependence on coupling constant,
however, right penal of in Fig.\ref{Fig.2} shows that EOS parameters for newly developed solution lies between $0$ and $\frac{1}{3}$,
Furthermore, Both attain maximum values at the center, which decrease when it moves towards the boundary of the stellar configuration.

The physical analysis of the solution has been presented in detail. The graphical representation of
energy and casuality conditions ensures the viability of the solution. However, adiabetic index does not meet the stability criterion
throughout inside the configuration which it inherits by its parent solution, however anisotropy aggravate the situation. The model successfully
represents the stellar configuration which may be crucial for gravitational collapse in future.

\section*{Acknowledgments}

``Authors thank the Higher Education Commission, Islamabad, Pakistan for its
financial support under the NRPU project with grant number
$\text{7851/Balochistan/NRPU/R\&D/HEC/2017}$''.

\end{document}